\documentclass[aps,pre,superscriptaddress,twocolumn]{revtex4-1}

\usepackage{mathtools}
\usepackage{color}

\newcommand{\be}{\begin{equation}}
\newcommand{\ee}{\end{equation}}
\newcommand{\la}{\langle}
\newcommand{\ra}{\rangle}
\newcommand{\Xfbm}{X_H}
\newcommand{\Xggbm}{X_{\beta,H}}
\newcommand{\odelta}{\overline{\delta^2}}
\newcommand{\EB}{E_{B}}
\newcommand{\EBfbm}{\EB^{\left(\rm fBm \right) }}
\newcommand{\EBggbm}{\EB^{\left(\rm ggBm \right)}}
\newcommand{\p}{\mathcal{P}}
\newcommand{\bx}{\mathbf{x}}

\begin{document}


\title{Fractional kinetics emerging from ergodicity breaking in random media}



\newcommand{\ikeraf}{\affiliation{Ikerbasque -- Basque Foundation for Science,
 Calle de Mar\'a D\'iaz de Haro 3, E-48013 Bilbao,  Basque Country, Spain}}
\newcommand{\istiaf}{\affiliation{ISTI-CNR, Istituto di Scienza e Tecnologie dell'Informazione "A. Faedo", Via Moruzzi 1, I-56124 Pisa, Italy}}
\newcommand{\bcamaf}{\affiliation{BCAM -- Basque Center for Applied Mathematics, Alameda de Mazarredo 14, E-48009 Bilbao, Basque Country, Spain}}
\newcommand{\icfoaf}{\affiliation{ICFO-Institut de Ci\`encies Fot\`oniques,\\The Barcelona Institute of Science and Technology, 08860 Castelldefels (Barcelona), Spain}}
\newcommand{\belgaf}{\affiliation{Department of Theoretical Physics, Belgorod National Research University, 14 Studencheskaya, 308015 Belgorod, Russia}}

\author{Daniel Molina--Garc\'ia}
\bcamaf

\author{Tuan Minh Pham}
\bcamaf
\belgaf

\author{Paolo~Paradisi}
\bcamaf
\istiaf

\author{Carlo Manzo}
\email[]{carlo.manzo@icfo.es}
\icfoaf

\author{Gianni Pagnini}
\email[]{gpagnini@bcamath.org}
\bcamaf
\ikeraf


\date{\today}

\begin{abstract}
We present a modelling approach for diffusion in a complex medium characterized by a random length scale. 
The resulting stochastic process shows subdiffusion with a behavior in qualitative agreement 
with single particle tracking experiments in living cells, such as ergodicity breaking, p-variation and aging. 
In particular, this approach recapitulates characteristic features previously described in part by 
the fractional Brownian motion and in part by the continuous-time random walk. 
Moreover, for a proper distribution of the length scale, a single parameter controls the ergodic-to-nonergodic transition and, 
remarkably, also drives the transition of the diffusion equation of the process from non-fractional to fractional, 
thus demonstrating that fractional kinetics emerges from ergodicity breaking. 

\bigskip
Published in: Physical Review E {\bf 94}, 052147 (2016) DOI: 10.1103/PhysRevE.94.052147
\end{abstract}


\maketitle
\section{Introduction}

Many processes in life sciences, soft condensed matter, geology, and ecology show a diffusive behaviour that cannot be modelled by classical methods. These phenomena are generally labeled with the term \emph{anomalous} diffusion in order to distinguish them from the \emph{normal} diffusion, where the adjective normal has the double aim of highlighting that: \emph{i})  
a Gaussian-based process is considered; and \emph{ii}) that it is a usual diffusion process with a linear growth in time of the particle displacement variance. The observation in nature of anomalous diffusion has been definitively established experimentally and several theoretical models have been proposed for the interpretation of such phenomenon~\cite{mercadier_etal-natphys-2009, barkai_etal-phystod-2012, metzler_etal-pccp-2014}. Among these theoretical efforts, the fractional calculus has emerged to be a successful tool for modelling a class of anomalous diffusion processes~\cite{delcastillonegrete_etal-prl-2005, metzler_etal-physrep-2000}. For this reason, anomalous diffusion governed by equations built on fractional derivatives is often also referred to as \emph{fractional} diffusion. Several stochastic approaches have been proposed in the literature to reproduce fractional kinetics~\cite{fulger_etal-per2008, pagnini-pa-2014, mura_etal-jpa-2008, gorenflo_etal-epjst-2011, magdziarz_etal-prl-2008}.

In the last decades, advances in fluorescence-based techniques such as single-particle-tracking (SPT) have allowed to precise characterization of the diffusion of molecules in biological systems~\cite{manzo_etal-ropp-2015}. In particular, the recording of long single-molecule trajectories has revealed that the occurrence of anomalous diffusion of some cellular components  in living cells is associated with ergodicity breaking (EB)~\cite{golding_etal-prl-2006, jeon_etal-prl-2011, weigel_etal-pnas-2011, tabei_etal-pnas-2013, manzo_etal-prx-2015}, i.e. the nonequivalence of time and ensemble averages~\cite{barkai_etal-phystod-2012, metzler_etal-pccp-2014}. Often EB and anomalous diffusion are concomitant with {\it aging},	i.e. the dependence of statistical quantities on the measurement time~\cite{burov_etal-pnas-2010}.

Besides the fundamental interest of nonergodic processes in statistical mechanics and its still unclear implications in cell biology, the occurrence of EB further embodies a valuable criterium for the selection of the underlying diffusive stochastic process. In this respect, comparative studies - involving the fractional Brownian motion (fBm)~\cite{deng_etal-pre-2009,magdziarz_etal-prl-2009}, the fractional Langevin equation~\cite{deng_etal-pre-2009} and the continuos time random walk (CTRW)~\cite{lubelski_etal-prl-2008,he_etal-prl-2008,magdziarz_etal-prl-2009} - have been conducted in order to determine which type of motion could possibly cause nonergodic anomalous diffusion. Among the mentioned theoretical frameworks, the fBm and the fractional Langevin motion are ergodic, with fBm displaying EB only in the ballistic limit~\cite{deng_etal-pre-2009}. 
On the other hand, the CTRW is nonergodic~\cite{lubelski_etal-prl-2008,he_etal-prl-2008} with the EB stemming from the nonstationary nature of the process when the distribution of waiting times has a power-law tail~\cite{lubelski_etal-prl-2008}. For this reason, the CTRW has been extensively used to model the occurrence of nonergodic diffusion and the waiting times have been associated to immobilization events caused by biochemical interactions~\cite{barkai_etal-phystod-2012, metzler_etal-pccp-2014}. 

However, due to the lack of nonergodic models alternative to the CTRW, the use of EB as a criterion to select the dynamic process has shown some limitations. An example is provided by the seminal work of Golding and Cox~\cite{golding_etal-prl-2006}. In this case, although the presence of EB favors CTRW as the model underlying the dynamics of RNA in cellular cytoplasm,  a moments-based criterion called p-variation~\cite{magdziarz_etal-prl-2009} seems to indicate a diffusion compatible with fBm. Similarly, other experiments also showed the simultaneous occurrence of EB and nonlinear scaling of the time-averaged mean-squared displacement, making necessary to hypothesise the coexistence of CTRW with other processes in order to theoretically model the observed features~\cite{jeon_etal-prl-2011, weigel_etal-pnas-2011, tabei_etal-pnas-2013}. 

In this paper, we provide a general framework in which EB emerges as a consequence of the heterogeneity (or randomness) of the system. The heterogeneity is described by the random nature of a characteristic property of the medium, such as a length scale  $\ell_\beta$, depending on a single parameter  $\beta$. Simple examples of this behavior are provided by a population of particles, each of them diffusing in a Brownian fashion but with a broad distribution of diffusion coefficients $\ell_\beta$. However, our conclusions do not depend on the type of motion performed by the particles. We also show that for any nontrivial choice of the distribution $\ell_\beta$, the parameter $\beta$ continuously drives the transition from ergodic to a nonergodic process. Notably, a fractional kinetics straightforwardly emerges from EB, and thus allows us to associate nonergodicity to a fractional equation.

For its generality, our approach constitutes a flexible tool to interpret the occurrence of EB in random media 
and in living cells without involving CTRW and subordination. From the biophysical point of view, it implies that EB can be generated by heterogeneity in the diffusion, without the need of particle trapping. In particular, we discuss how our model can resolve the controversy on the interpretation of Golding and Cox experiments~\cite{golding_etal-prl-2006, he_etal-prl-2008, neusius_etal-pre-2009, magdziarz_etal-prl-2009} by considering the fBm in a heterogeneous medium. Such a model allows one to simultaneously obtain the apparently contradictory features observed in Golding and Cox experiments, i.e. the monotonically increasing p-variation test typical of the fBm together with the EB parameter of the CTRW.

Finally, we show that our formulation can be further generalized by considering a nonstationary length scale $\ell_\beta=\ell_\beta(t)$ and thus including the occurrence of aging. 

\section{Ergodicity breaking from diffusion in a random medium}
In our model, we consider a stochastic process defined as:
\begin{equation*}
X(t)=\ell_\beta \, X_{gen}(t) \, ,
\end{equation*}
describing a population of particles diffusing according to a generic ergodic Gaussian process $X_{gen}(t)$ in a 
complex random medium.  The medium properties are independent of the diffusing particles and its randomness is described by a random characteristic quantity - such as a length scale $\ell_\beta$ - with distribution depending on the parameter $\beta$.  The role of $\beta$ thus consists in tuning the degree of randomness of the medium 
by modulating the distribution of the length scale. For for the case in which $X_{gen}(t)$ represents a random walk, $\ell_\beta$ corresponds to a distribution of diffusion coefficients.


Although the following conclusions hold for every ergodic Gaussian stochastic process, for the sake of simplicity from now on we will consider the fBm $\Xfbm(t)$, an ergodic non-Markovian Gaussian process characterized by the covariance matrix:
\be
\gamma_H(t,s)=t^{2H} + s^{2H} - |t-s|^{2H} \,,
\label{gammaH}
\ee
where $0 < H < 1$ is the Hurst exponent, and the variance results to be $\la \Xfbm^2 \ra=2 \, t^{2H}$.

Therefore, we investigate the following diffusion process $X(t)$ in a random medium:
\be
X(t)=\ell_\beta \, \Xfbm(t) \,.
\label{ggBm0}
\ee

In order to study the dynamics of the process, we first consider 
the time-averaged mean-square displacement 
\cite{lubelski_etal-prl-2008,he_etal-prl-2008,deng_etal-pre-2009}
\be
\odelta(T) = \frac{\int_0^{T-\Delta} \, [X(\xi+\Delta)-X(\xi)]^2 \, d\xi}{T-\Delta} \,,
\ee
where $\Delta$ is the timelag and $T$ the measurement time. The time-averaged mean-square displacement describes the time dependence of the second moment of the particle's position and it is often used to classify the diffusion mode.  For the pure Brownian motion ($2H=1$), $\odelta(T)$ shows a linear growth with $\Delta$, whereas for the fBm shows a power-law behaviour $\sim\Delta^{2H}$, i.e. anomalous diffusion.
The effect of the random length scale is preserved in the calculation of $\odelta(T)$. For the particular case $2H=1$ in which the process $\Xfbm(t)$ in (\ref{ggBm0}) corresponds to the pure Brownian motion, the random length scale is proportional to the diffusion coefficient. Consequently, as shown in Fig.~\ref{TAMSDggBm}, time averages such as $\odelta(T)$ remain random variables and thus irreproducible~\cite{burov_etal-pccp-2011}, causing ergodicity breaking (EB). 
This effect can be estimated through the calculation of the EB parameter $\EB(T)$~\cite{he_etal-prl-2008,deng_etal-pre-2009}. Let $\la \cdot \ra$ represent the ensemble averaging,
then 
\be
\EB(T)=\frac{\la [\odelta(T)]^2 \ra}{\la \odelta(T) \ra^2} - 1 
\label{EB}
\ee
is calculated in the large $T$ limit and tends to $0$ when the process is ergodic~\cite{deng_etal-pre-2009}. \\

With a fixed and non-random length scale, e.g. $\ell_\beta=1$, for the stochastic process $X(t)$ defined in (\ref{ggBm0}), 
we obtain \cite{deng_etal-pre-2009}
\be
\EB^{( \ell_\beta=1 ) }(T)=\EBfbm(T) \xrightarrow{T \to \infty} 0 \,.
\ee 
In contrast, if $\ell_\beta$ is a random variable, for $X(t)$ it holds that
\be
\EB^{( \ell_\beta )}(T)=\frac{\la \ell_\beta^4 \ra}{\la \ell_\beta^2 \ra^2}\left[\EBfbm(T)+1\right]-1  
\xrightarrow{T \to \infty} \frac{\la \ell_\beta^4 \ra}{\la \ell_\beta^2 \ra^2}-1 \,.
\label{EBlimit0}
\ee 

The condition $\la \ell_\beta^4 \ra > \la \ell_\beta^2 \ra^2$ is met in general for any distribution 
as a consequence of the inequality $K \geq S^2 +1$ \cite{kendall_stuart-1977}, 
where $K$ and $S$ are the kurtosis and the skewness respectively, 
and in particular for any unilateral nonincreasing density it holds $K \geq 9/5$ from the Gauss--Winckler inequality \cite{kendall_stuart-1977}.
The limiting case $\la \ell_\beta^4 \ra = \la \ell_\beta^2 \ra^2$ is met when the distribution of the length scale is 
the Bernoulli distribution with equal success probability for values $0$ and $1$ or it is the Dirac-delta distribution 
$\delta(\ell_\beta-1)$; 
therefore the process is nonergodic for every nontrivial choice of $\ell_\beta$. 
Although these conclusions might look somehow trivial, they show how a complex medium - through a random distribution of the length scale - might produce nonergodic behavior into an ergodic Gaussian stochastic process, including the pure Brownian motion, only by introducing heterogeneity~\cite{lubelski_etal-prl-2008}. 


\section{Ergodicity breaking and the fractional kinetics}

In the previous section we have shown that, since $\ell_\beta$ is an independent random variable, EB can occur as the sole consequence of the randomness of the medium in which diffusion takes place (\ref{EBlimit0}) and independently of the chosen ergodic Gaussian stochastic process.

In the following, we will focus our attention on the stochastic process $X(t)$ as defined in (\ref{ggBm0}). This process has already been studied in a specific characterization
named {\it generalized grey Brownian motion} (ggBm) \cite{mura_etal-jpa-2008,mura_etal-pa-2008,mura_etal-itsf-2009}.
As a matter of fact, the ggBm trajectory $\Xggbm(t)$ is obtained by setting $\ell_\beta=\sqrt{\Lambda_\beta}$, i.e.
\be
\Xggbm(t) = \sqrt{\Lambda_\beta} \, \Xfbm(t) \,, 
\label{ggBm}
\ee
where the positive random variable $\Lambda_\beta$ is distributed according to 
the one-side M-Wright/Mainardi function $M_\beta(\lambda)$, with $\lambda \ge 0$ and $0 < \beta < 1$, defined as \cite{mainardi_etal-ijde-2010,pagnini-fcaa-2013}
\be
M_\beta(\lambda)=\sum_{k=0}^\infty \frac{(-1)^k}{k!} \frac{\lambda^k}{\Gamma[-\beta k + (1-\beta)]} \,.
\ee
The case of a non-random length scale, i.e. $\Lambda_\beta=1$, 
is straightforwardly recovered in the limit $\beta \to 1$
since it holds $M_1(\lambda)=\delta(\lambda-1)$.
The ggBm is a rather general model and includes as special cases the Brownian motion ($\beta=2H=1$),
the fBm ($\beta=1$) and the grey Brownian motion ($\beta=2H$).

It is well known that the probability density function of $\Xggbm(t)$ is \cite{mura_etal-jpa-2008} 
\begin{widetext}
\be
\p(\bx;\gamma_H)=\frac{1}{\sqrt{(2\pi\lambda)^n \, {\rm det} \, \gamma_H}} 
\int_0^\infty \exp\left\{-\frac{1}{2\lambda} \, \bx^T \gamma_H^{-1} \bx\right\} \, M_\beta(\lambda) \, d\lambda \,,
\label{ggBmpdfn}
\ee
\end{widetext}
where $\bx=(x_1,\dots,x_n)$ and $\gamma_H=\gamma_H(t_i,t_j)$, $i,j=1,\dots,n$, 
is the covariance matrix of the fBm defined in (\ref{gammaH}).
Therefore, by the Mellin transform of $M_\beta(\lambda)$ \cite{mainardi_etal-fcaa-2003},
i.e. $\int_0^\infty \lambda^{s-1} M_\beta(\lambda) \, d\lambda=\Gamma[1+(s-1)]/\Gamma[1+\beta(s-1)]$,
with $s > 0$, the covariance matrix of the ggBm can be obtained as 
\cite{mura_etal-jpa-2008,mura_etal-itsf-2009}
\be
\gamma_{\beta,H}(t,s)=\frac{1}{\Gamma(1+\beta)}(t^{2H} + s^{2H} - |t-s|^{2H}) \,.
\label{ggBmcovariance}
\ee
The one-point one-time density function can be derived from (\ref{ggBmpdfn}) and becomes 
\begin{eqnarray}
\hspace{-0.5truecm}
\p(x;t) &=& \frac{1}{\sqrt{4 \pi \lambda \, t^{2H}}}\int_0^\infty \exp\left\{-\frac{x^2}{4 \lambda \, t^{2H}}\right\}
\, M_\beta(\lambda) \, d\lambda \\
&=& \frac{1}{2 \, t^H} \, M_{\beta/2}\left(\frac{|x|}{t^H}\right) \,,
\label{fbmpdf}
\end{eqnarray}
where it emerges that the shape of probability density function of displacements is affected by the medium, here
represented by $M_\beta(\lambda)$. 
In terms of the H-function the density function $\p(x;t)$ reads \cite{mainardi_etal-jcam-2005,mainardi_etal-jcam-2007}
\be
\p(x;t)=\frac{1}{2 \, t^H} \, 
H^{10}_{01}\left[\frac{|x|}{t^H} \left|
\begin{array}{ccc}
- & \,; & (1-\beta/2,\beta/2)\\
\\
(0,1) & \,; & -
\end{array}
\right.
\right] \,,
\ee
and the asymptotic decay is $M_{\beta/2}(|x| \to \infty) \sim |x|^{\frac{c}{2}(\beta-1)} \, {\rm e}^{-b|x|^c}$, 
with $b=\frac{2^{1-c}}{c} \, \beta^{\beta c/2}$ and $c=\frac{2}{2-\beta}$~\cite{mainardi_etal-fcaa-2001,mainardi_etal-jcam-2007}.
From (\ref{ggBmcovariance}) the variance turns out to be
\be
\la \Xggbm^2 \ra = \frac{2}{\Gamma(1+\beta)} \, t^{2H} \,,
\label{fbmvariance}
\ee
showing that the presence of the medium does not affect the power law growth of the particle displacement variance over time. 
It is noteworthy to observe that the ggBm shows both subdiffusion, $0 < H < 1/2$, and superdiffusion, $1/2 < H < 1$.
Moreover, a remarkable case is represented by $H=1/2$ in which the particle displacement variance results to be linear in time, see (\ref{fbmvariance}),
but the density function is not Gaussian according to (\ref{fbmpdf}). The Gaussian density is obtained from (\ref{fbmpdf}) as a special case when $\beta=1$.

The evolution equation for $\p(x;t)$ is given by
\be
\frac{\partial \p}{\partial t} = 
\frac{2H}{\beta} t^{2H-1} \, \mathcal{D}_{2H/\beta}^{\beta-1,1-\beta} \frac{\partial^2 \p}{\partial x^2} \,,
\label{EKeq}
\ee
where $\mathcal{D}_{\eta}^{\xi,\mu}$ is the Erd\'elyi--Kober fractional derivative 
with respect to $t$
and then the process is also referred to as {\it Erd\'elyi--Kober fractional diffusion} \cite{pagnini-fcaa-2012}.
Special cases of Eq. (\ref{EKeq}) are: the classical diffusion ($\beta=2H=1$),
the fBm master equation ($\beta=1$) and the time-fractional diffusion equation ($\beta=2H$).
A similar approach can be developed in the framework of the space-time fractional diffusion equation,
which includes all its special cases \cite{pagnini_etal-fcaa-2016}.

We would like to remark that the fractional kinetics, i.e. $\beta \ne 1$, 
emerges directly from the EB due to the randomness of $\ell_\beta=\sqrt{\Lambda_\beta}$
since $M_{\beta \ne 1}(\lambda) \ne \delta(\lambda-1)$. Moreover, the fractional order related to $\beta$ can be experimentally computed
by means of the long-time limit of the EB parameter. In fact, for large $T$,
from (\ref{EBlimit0}) and $\ell_\beta=\sqrt{\Lambda_\beta}$ the EB parameter $\EBggbm(T)$ then becomes
\be
\EBggbm(T) \xrightarrow{T \to \infty}
\frac{\la \Lambda_\beta^2 \ra}{\la \Lambda_\beta \ra^2} - 1=
\beta \, \frac{\Gamma(\beta)\Gamma(\beta)}{\Gamma(2\beta)}-1 \,,
\label{EBlimitggBm}
\ee
where again the Mellin transform of $M_\beta(\lambda)$ \cite{mainardi_etal-fcaa-2003}
has been used to compute $\la \Lambda_\beta^2 \ra$ and $\la \Lambda_\beta \ra$.

In summary, the existence of a random length scale turns an ergodic process into a nonergodic one
without the need to introduce an alternative stochastic process. When this transition occurs continuously with respect to a parameter $\beta$,
the distribution of the length scale can be related to the M-Wright/Mainardi function
and the resulting stochastic process is driven by a fractional diffusion equation.

Therefore, 
the present formulation provides a foundation of fractional kinetics on the basis
of the appearance of the EB. In other words, fractional kinetics can be considered as stemming from the EB due to
the heterogeneity of the medium in which the diffusion takes place.
In order to support this physical foundation argument, 
we remark that from the proposed ggBm (\ref{ggBm}) the evolution of the particle density function is governed by a fractional diffusion equation
also in the special case $H=1/2$,  see (\ref{EKeq}), with $X_H(t)$ performing the classical Brownian motion and 
the particles displaying a variance with a linear growth in time (\ref{fbmvariance}). 
\section{Relation with experiments}

Advances in biophysical techniques, such as SPT, have allowed researchers to detail the motion of single molecules and have revealed very complex diffusion patterns in living-cells~\cite{manzo_etal-ropp-2015}. In particular, the analysis of these experiments has shown that several biological systems display nonergodic behavior as a consequence of interactions occurring in heterogeneous cellular environments~\cite{golding_etal-prl-2006, jeon_etal-prl-2011, weigel_etal-pnas-2011, tabei_etal-pnas-2013, manzo_etal-prx-2015}. Such nonergodic behavior has often been connected with the occurrence of anomalous (sub)diffusion. The occurrence of EB has been mainly identified through the nonequivalence of time and ensemble averages and by the calculation of the EB parameter (\ref{EB}) \cite{he_etal-prl-2008,deng_etal-pre-2009}.  Owing to the importance of molecular transport for the cellular function, theoretical efforts have been devoted to understand the physical mechanism behind EB in biology.  Several stochastic models presenting nonstationary (and thus nonergodic) (sub)diffusion have been proposed~\cite{metzler_etal-pccp-2014}. Among these models, the most popular has definitively been the CTRW \cite{montroll_etal-jmp-1965, scher_etal-prb-1975, lubelski_etal-prl-2008,he_etal-prl-2008,magdziarz_etal-prl-2009} which has been extensively used to model nonergodic subdiffusion in living cells ~\cite{jeon_etal-prl-2011, weigel_etal-pnas-2011, tabei_etal-pnas-2013}. The CTRW has allowed association of the nonergodic behavior with the occurrence of particle immobilization with a heavy-tailed distribution of trapping times~\cite{bel_etal-prl-2005}.

However, among the experimental evidences of EB in biological systems, not all the observed features could be directly addressed within the framework of CTRW alone. For example, Refs.~\cite{golding_etal-prl-2006, weigel_etal-pnas-2011, tabei_etal-pnas-2013} showed subdiffusive scaling of the time-averaged mean-square displacement obtained for single trajectories, making necessary the
postulation of the coexistence of CTRW with other sources of subdiffusion, i.e. the fBm~\cite{tabei_etal-pnas-2013} or a fractal processes~\cite{weigel_etal-pnas-2011}, in order to properly interpret the results. In addition, some experiments did not show the occurrence of inherent features of CTRW, such as {\it aging}~\cite{bronstein_etal-prl-2009} or immobilization~\cite{manzo_etal-prx-2015}.

In order to determine the physical scenario behind the subdiffusive EB, a number of diagnostic tools have been proposed~\cite{meroz_etal-phyrep-2015}. Among these, a valid criterion for selection of stochastic processes  
is represented by the so-called p-variation test~\cite{magdziarz_etal-prl-2009}. The test is based on the calculation of the quantity
\be
V^{(p)}(t)=\lim_{n \to \infty} V^{(p)}_n(t) \,,
\label{Vp}
\ee
where for $t \in [0,T]$
\be
\hspace{-0.23truecm}
V^{(p)}_n(t)=\sum_{j=0}^{2^n-1}\left|
X\left(\frac{(j+1)T}{2^n} \wedge t\right)-X\left(\frac{jT}{2^n} \wedge t\right)\right|^p \,, 
\label{Vpn}
\ee
with $a \wedge b = \min\{a,b\}$, and allows the CTRW-like models and the fBm to be distinguished, even on the single trajectory level~\cite{meroz_etal-phyrep-2015}.

In spite of the efforts in developing tests and methods to distinguish between different stochastic models, contradictory indications still prevent the unambiguous determination of the physical mechanism behind EB in biological samples. An example is provided by what is probably the first evidence of EB in living cells, i.e. the experiments describing the motion of individual mRNA molecules inside living {\it E. coli} cells presented in the seminal paper by Golding and Cox~\cite{golding_etal-prl-2006}.  In this case, in order to explain the occurrence of EB as evidenced by the large scattering of single-trajectory $\odelta$ curves and a non-zero EB parameter, the CTRW was proposed in Refs.~\cite{he_etal-prl-2008,neusius_etal-pre-2009} to model this dataset. However, in order to account for the subdiffusive behavior of the time-averaged mean-square displacement, the authors of both works proposed the coexistence of CTRW with some degree of spatial confinement producing the power law behavior of $\odelta$~\cite{he_etal-prl-2008,neusius_etal-pre-2009}. 
But the application of the p-variation test to the same dataset \cite{golding_etal-prl-2006} showed that the subdiffusion is unlikely to originate from the CTRW, whereas the data are compatible with fBm~\cite{magdziarz_etal-prl-2009}.

In this scenario, the general stochastic process presented in this work in (\ref{ggBm0}) provides a plausible framework to describe the subdiffusive nonergodic behavior observed in Ref.~\cite{golding_etal-prl-2006}. 
The introduction of a random length scale associate to a random medium allows to describe the complexity of the cytoskeletal environment and reproduce the scatter of time-average mean-square displacement observed at the single trajectory level. 
This observation is quantitatively translated by the calculation of the EB parameter.  
As a matter of fact, equation (\ref{EBlimitggBm}) shows that the EB parameter of the specific process described in (\ref{ggBm}) is identical to the one obtained for  a CTRW with a power law distribution of waiting times, i.e. $\psi(\tau) \propto \tau^{-(1+\beta)}$, and infinite average sojourn time~\cite{he_etal-prl-2008,neusius_etal-pre-2009}, independently of the ergodic Gaussian process used to model diffusion.
In addition, the flexibility of our method allows us to choose the fBm to model single particle diffusion  (\ref{ggBm})  and thus reproduce the subdiffusion in $\odelta$ and maintain the same p-variation behaviour of the fBm $V_{ggBm}^{(p)}(t)=\Lambda_\beta^{p/2} \, V^{(p)}_{f\!Bm}(t)$, while preserving the same degree of EB observed for CTRW-like models (Fig.~\ref{datacompar}).

\section{Aging}

An interesting feature emerging from some single-particle tracking experiments of cellular components~\cite{weigel_etal-pnas-2011, tabei_etal-pnas-2013, manzo_etal-prx-2015} is the occurrence of {\it aging}, i.e. the dependence of statistical quantities - such as the time and ensemble averaged mean-square displacement  $\la \odelta(T)\ra$ - on the measurement time, as a consequence of the presence of nonstationarity in the diffusive mechanism~\cite{burov_etal-pnas-2010}.
Besides living cells, aging has been observed for many complex systems such as blinking nanocrystals~\cite{margolin_etal-jcp-2004,paradisi_etal-aipcp-2005,bianco_etal-jcp-2005}, spin glasses~\cite{jonason_etal-prl-1998} and colloidal suspension~\cite{abou_etal-pre-2001}.
Since aging can characterize long-term memory ~\cite{akin_pa06}, it can be used as a statistical indicator of complexity and thus exploited to discriminate among different modelling approaches ~\cite{paradisi_cejp09,akin_jsmte09}. Furthermore, aging has been shown to be associated with weak ergodicity breaking~\cite{bouchaud-jpf-1992, schulz_etal-prl-2013}, i.e. a situation in which the time needed to explore a system phase space is infinite but the phase space can not be divided into mutually inaccessible regions~\cite{bouchaud-jpf-1992}.

\begin{figure}
\includegraphics[width=\columnwidth]{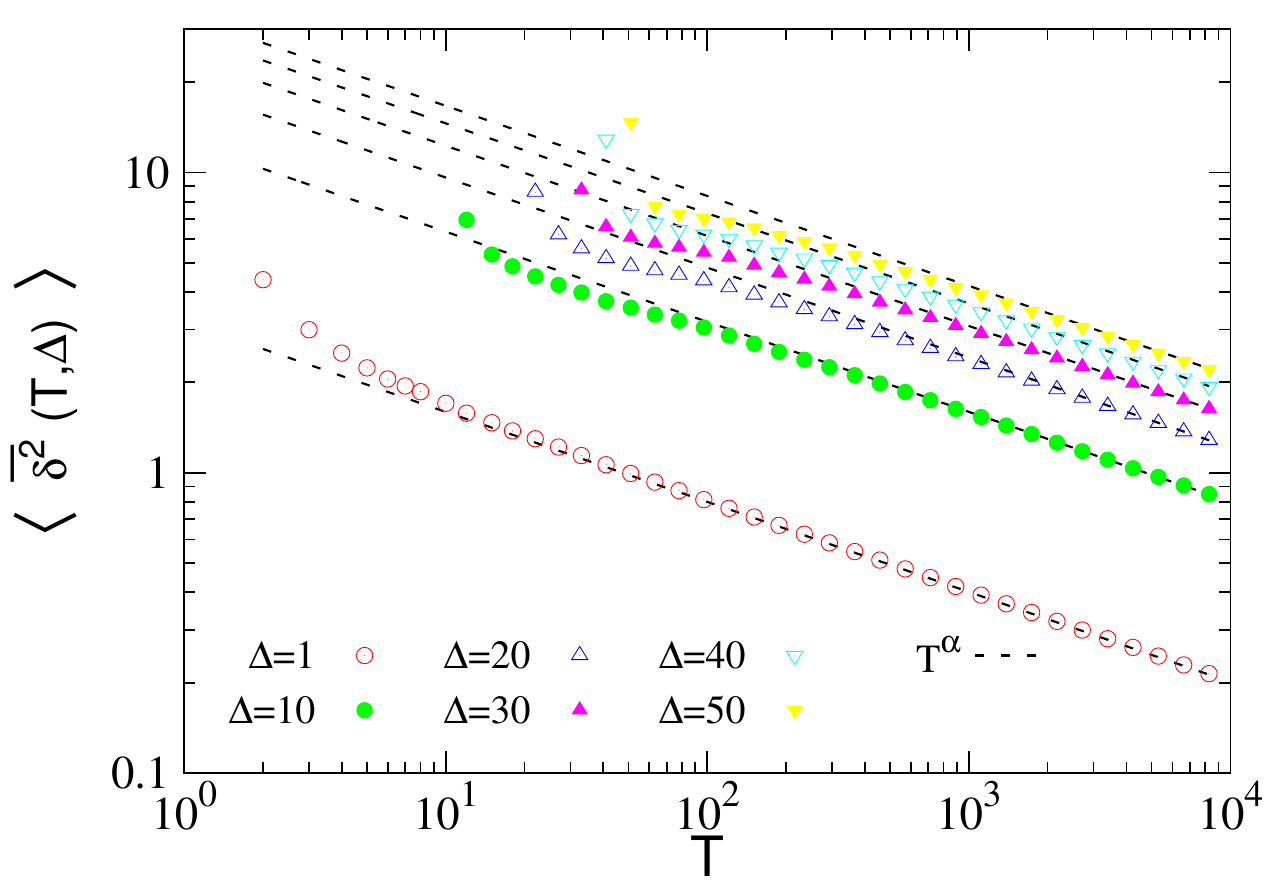}%
\caption{
Plot of the time and ensemble averaged mean-square displacement $\la \odelta(T)\ra$ at various timelags $\Delta$ and as a fucntion of the measurement time $T$ for 
the process (\ref{nonstatprocess}) with $\beta=H=0.3$ and $\alpha=-0.3$. The curves asymptotically show a power law decay $T^\alpha$ (dashed lines) demonstrating the presence of aging in the process.}
\label{aging}
\end{figure}

Our theoretical formulation allows to reproduce aging by the extension to the case of a nonstationary random medium
$\ell_\beta=\ell_\beta(t)$~\cite{molina_etal-nolasc15-2015}. 
The stochastic process results to be defined as: 
\be
X_{\alpha,\beta,H}(t)=\sqrt{t^\alpha \Lambda_\beta} \, X_H(t) \,,
\label{nonstatprocess}
\ee
where $\Lambda_\beta$ and $X_H(t)$, with $0<H<1$, have the same meaning as in equation (\ref{ggBm}). In this case, the increments of $X_{\alpha,\beta,H}(t)$ are nonstationary, in contrast to the process defined in (\ref{ggBm}), which is recovered as a particular case for $\alpha=0$. The parameter $\alpha$ is constrained by the physical requirement that the process is diffusive, meaning that the particle displacement variance must grow in time. Since the variance of the process is given by
\be
\la X^2_{\alpha,\beta,H} \ra =\la \Lambda_\beta \ra \, t^{\alpha + 2H} \,,
\label{XabH}
\ee 
the latter condition can be expressed as $\alpha > -2H$. It can be shown \cite{molina_etal-nolasc15-2015} 
that the time- and ensemble-averaged mean-square displacement then is $\la \overline{\delta^2(T)} \ra \simeq \Delta^{2H} T^\alpha$ (Fig.~\ref{aging}). 

\begin{figure}
\includegraphics[width=\columnwidth]{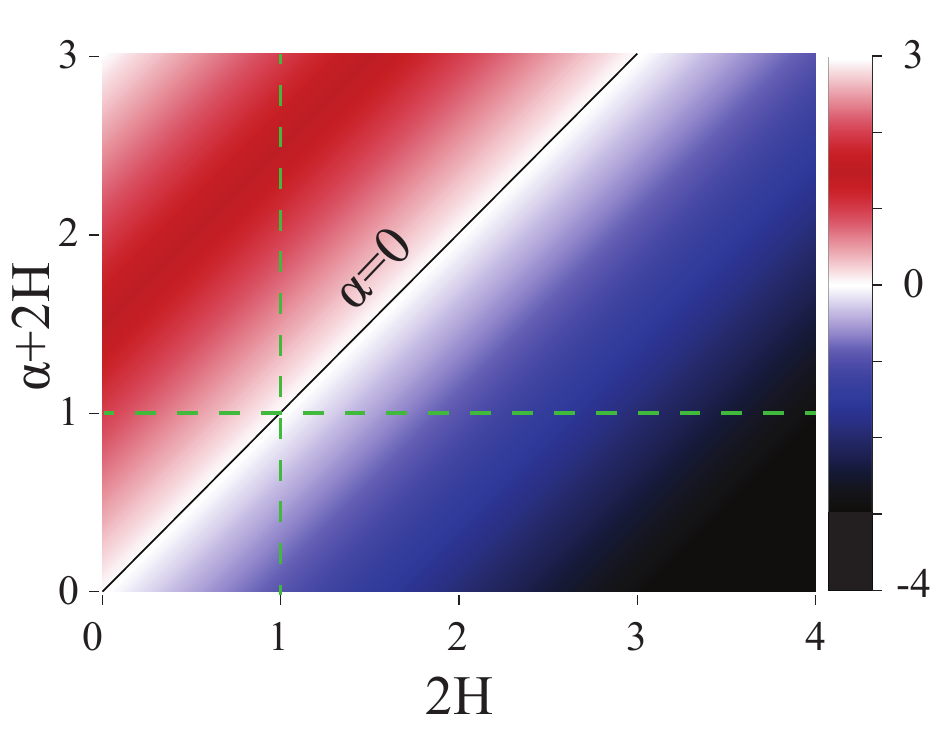}%
\caption{
Contour plot of the aging exponent $\alpha$ as a function of the exponents controlling the power law growth of the time- ($2H$) and ensemble-averaged  ($\alpha+2H$) mean-squared displacement for the process (\ref{nonstatprocess}). The continuous black line corresponds to the absence of aging ($\alpha=0$). Dashed green lines separate sub- and super-diffusive regions, characterized by exponents values $<1$ and $>1$, respectively. }
\label{alpha_vs_H}
\end{figure}

It is interesting to note that our formulation shows properties that were not recapitulated by any of the models for nonergodic diffusion previously presented in literature~\cite{metzler_etal-pccp-2014}.  First, the exponents controlling the power law behavior of $\odelta$, $\la X^2_{\alpha,\beta,H} \ra$ and $\la \odelta(T)\ra$  depend on two parameters, $\alpha$ and $H$. 
As such they can thus be independently tuned to reproduce any different scaling of the two curves, in contrast to the other models~\cite{metzler_etal-pccp-2014}. 
In particular, our model show that the time- and ensemble-averaged mean squared displacements can have marked different behavior, 
for example with one showing subdiffusivity while the other showing superdiffusivity. In addition, the aging can shows positive or negative exponent depending on the relative magnitude of the exponents controlling the growth of the  time- and ensemble-averaged mean squared displacement (Fig. \ref{alpha_vs_H}). 

Moreover, we highlight that the aging can be obtained even in the case in which the time-averaged mean squared displacement $\overline{\delta^2}$ or the ensemble-averaged mean squared displacement $\la X^2_{\alpha,\beta,H}\ra$ show Brownian behaviour, i.e. when $2H = 1$ or $\alpha=1$, respectively. It is interesting to note that in the case $2H = 1$ we recover the same relationship between the exponent of the ensemble-averaged mean squared displacement ($\alpha+1$) and the time-ensemble-averaged mean squared displacement obtained for other models, such as the CTRW~\cite{lubelski_etal-prl-2008}, the scaled Brownian motion~\cite{jeon_etal-pccp-2014}, the quenched trap~\cite{miyaguchi_etal-pre-2011} and the patch model~\cite{massignan_etal-prl-2014}. Moreover, the calculation of the EB parameter (\ref{EB}) for the process (\ref{nonstatprocess}) shows that even in the presence of aging ($\alpha \ne 0$) the value of the EB parameter is identical to the one obtained for a CTRW with infinite average sojourn time and power law distribution of waiting times~\cite{molina_etal-nolasc15-2015}. 

\section{Conclusions}

We have demostrated that an ergodic Gaussian process occurring in a heterogeneous medium characterized by a random length scale can be turned into nonergodic without altering the properties of the Gaussian process itself.  We showed that for any nontrivial choice of the distribution of the length scale, the transition from ergodicity to nonergodicity can be continuoulsy tuned by means of a parameter $\beta$. In these cases, the distribution of the length scale can be related to the M-Wright/Mainardi function and the resulting stochastic process is controlled by a fractional diffusion equation.

\begin{figure}
\includegraphics[width=\columnwidth]{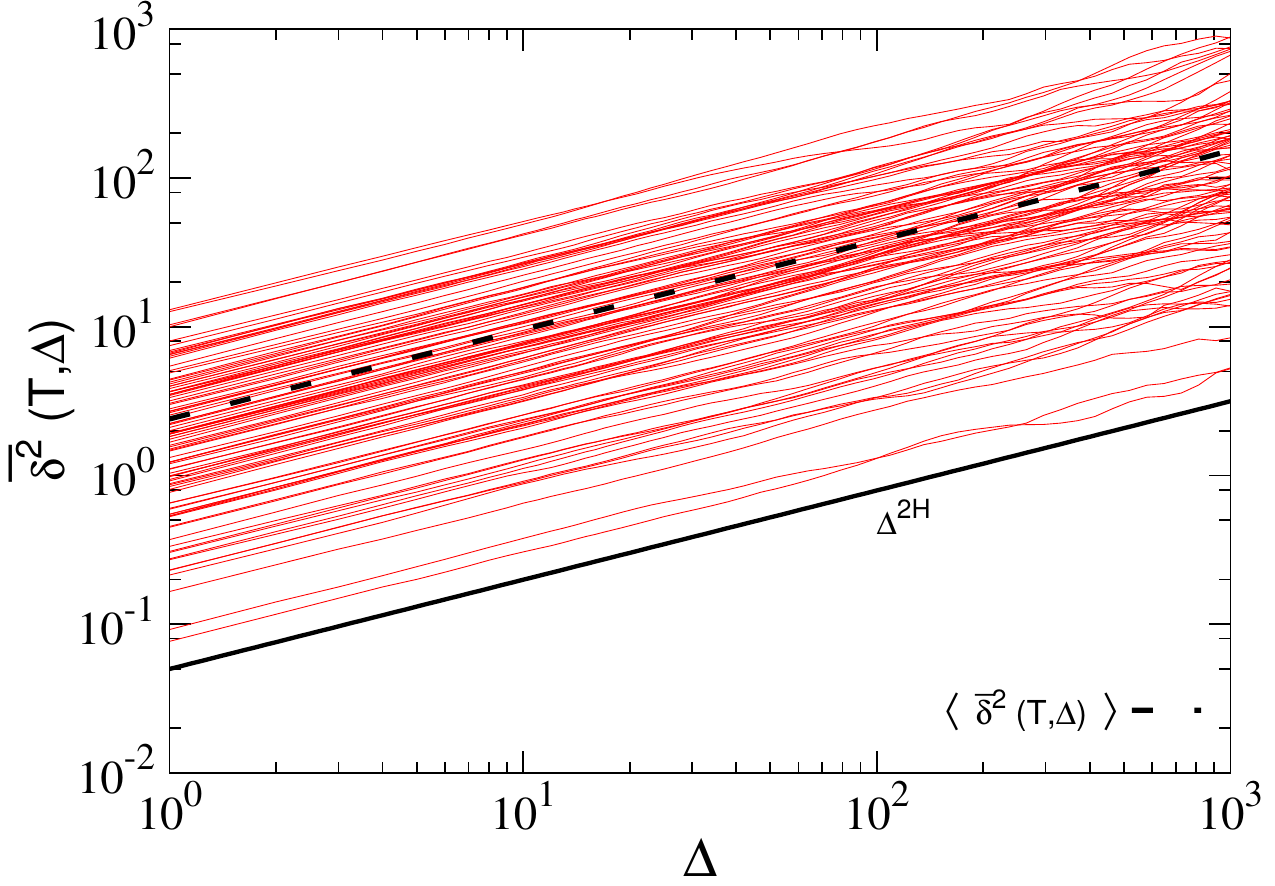}%
\caption{Time-averaged mean-square displacement $\odelta$ as a function of the timelag $\Delta$ calculated for several trajectories (thin red lines) performing the fBm in a random medium, according to the ggBm (\ref{ggBm}) with $\beta=H=0.3$ and $T=10^4$. Dashed line corresponds to the time and ensemble averaged mean-square displacement. Continuous thick line is a guide to the eye.}
\label{TAMSDggBm}
\end{figure}

\begin{figure}
\includegraphics[width=\columnwidth]{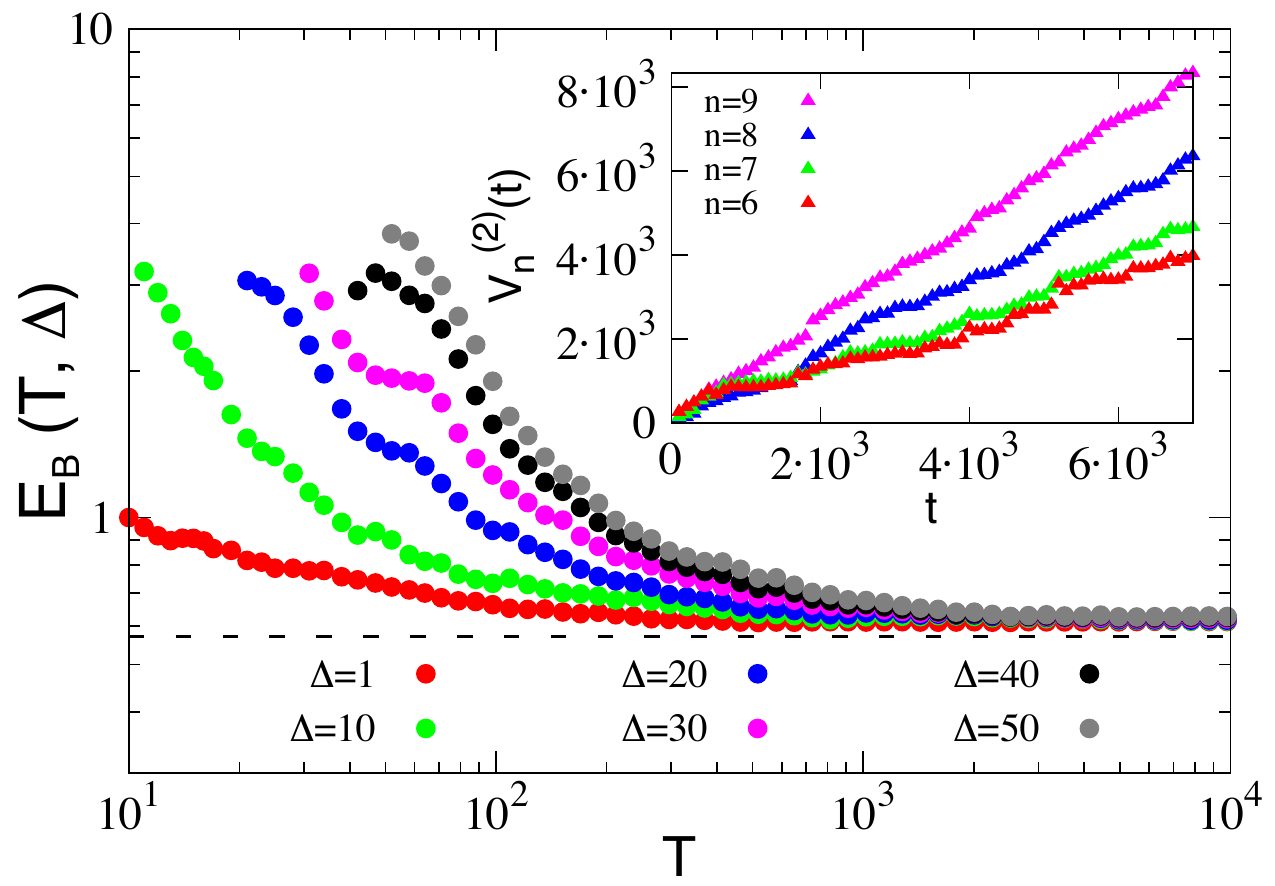}%
\caption{
Plot of $E_{B}(T)$ for the stochastic process (\ref{ggBm}) with $\beta=0.5$ and $H=0.3$ at various timelags $\Delta$ as a function of the measurement time $T$. Larger $\Delta$ produces an increase of $E_{B}(T)$ at short time $T$.
The $E_{B}(T)$ values at large time $T$ show are in agreement with the theoretical expectation (\ref{EBlimitggBm})  (dashed line). (Inset) Results of the the p-variation test with $p = 2$ for the stochastic process  (\ref{ggBm})  with $\beta = 0.5$ and $H = 0.3$, showing the same trend as the pure fBm.}
\label{datacompar}
\end{figure}

These conclusions are valid for any ergodic Gaussian process. Therefore, the generality of our formulation posits it as a flexible tool for the interpretation of heterogeneous and/or nonergodic diffusion in disordered systems, such as the many examples of subdiffusion recently observed in living cells~\cite{golding_etal-prl-2006, bronstein_etal-prl-2009, jeon_etal-prl-2011, weigel_etal-pnas-2011, tabei_etal-pnas-2013, manzo_etal-prx-2015}. Notably, our formulation includes the possibility to model the simultaneous occurrence of subdiffusion (as well as any other types of motion) at the single particle level (Fig. \ref{TAMSDggBm}) and EB (Fig. \ref{datacompar}), a feature observed in many experimental reports~\cite{golding_etal-prl-2006, jeon_etal-prl-2011, weigel_etal-pnas-2011, tabei_etal-pnas-2013}. This is in contrast with other nonergodic models, such as the CTRW, predicting a linear scaling of the time-averaged mean-squared displacement. Therefore, the data could not be satisfactorily interpreted by the CTRW alone and needed to include an additional source of subdiffusion together with CTRW models~\cite{weigel_etal-pnas-2011, tabei_etal-pnas-2013}.

In particular, we showed that our framework offers an interpretation of the data of Golding and Cox~\cite{golding_etal-prl-2006} on the basis of a fBm in a heterogeneous medium. The stochastic process (\ref{ggBm}) allows to capture both the subdiffusivity in the time-averaged mean-squared displacement, the monotonic temporal growing of the p-variations test (as for the fBm), as well as the EB parameter value of the CTRW. Therefore, our model allows to reproduce all the features observed experimentally and thus solve the disagreement about the underlying stochastic process. 
 
Furthermore, we show that by introducing a nonstationary random medium (\ref{nonstatprocess}), our model can be extended to include the occurrence of aging, a feature often associated to EB in living systems~\cite{weigel_etal-pnas-2011,manzo_etal-prx-2015}. As such, we consider that our general approach could contribute to investigate the occurrence of EB and anomalous diffusion in life sciences as well as many other fields, and help to elucidate their effects and implications.

\newpage
\begin{acknowledgments}
This research is supported by the Basque Government through the BERC 2014-2017 program and 
by the Spanish Ministry of Economy and Competitiveness MINECO: 
BCAM Severo Ochoa accreditation SEV-2013-0323 and Grant No. MTM2016-76016-R. 
C.M. acknowledges support from the Spanish Ministry of Economy and Competitiveness 
(``Severo Ochoa'' Programme for Centres of Excellence in R\&D (SEV-2015-0522) and FIS2014-56107-R), 
Fundaci\'o Privada CELLEX (Barcelona), HFSP (GA RGP0027/2012), and LaserLab Europe 4 (GA 654148).
P.P. acknowledges support from the Bizkaia Talent Organization 
(2015 Financial Aid Programme for Researchers, Project No. AYD-000-252) 
and European Commission through the COFUND program.
\end{acknowledgments}


%

\newpage

\end{document}